\documentclass{article}
\usepackage{spconf,amsmath,graphicx,hyperref}

\usepackage{fix-cm}
\usepackage{amsmath}
\usepackage{booktabs}
\usepackage{graphicx}
\usepackage{multirow}
\usepackage{xcolor}
\usepackage{enumitem}
\usepackage{amssymb}
\usepackage[fontsize=9pt]{fontsize}

\usepackage[numbers,sort&compress]{natbib}
\let\OLDthebibliography\thebibliography
\renewcommand\thebibliography[1]{
  \OLDthebibliography{#1}
  \setlength{\parskip}{4pt}
  \setlength{\itemsep}{-0.5ex}
}

\title{Mitigating Stethoscope-Induced Shortcuts in Respiratory Sound Classification under Federated Domain Generalization with Causality-Inspired Interventions}

\name{Heejoon Koo$^{1}$$\thanks{\hspace{-1em}$^{\dagger}$ denotes corresponding author.}$, Yoon Tae Kim$^{2}$, Miika Toikkanen$^{2}$, June-Woo Kim$^{3\dagger}$}
\address{$^{1}$University of Illinois Urbana-Champaign, USA, $^{2}$RSC LAB, MODULABS, Korea, $^{3}$Wonkwang University, Korea}

\begin{document}

\maketitle

\begin{abstract}
AI-driven respiratory sound classification (RSC) is promising for automated pulmonary disease detection, yet multi-site deployment is hindered by inter-stethoscope variability. We introduce a federated domain generalization (FedDG) formulation for RSC in which clients hold recordings from different stethoscopes and the model is evaluated on an unseen device. Our empirical analysis shows that stethoscope-induced style and disease-relevant content are partially entangled, making deterministic style removal unreliable. In response, we propose BTS-CAFE, a framework combining (i) causality-inspired device-style interventions with constraints designed to limit content distortion, (ii) counterfactual metadata augmentation to relieve device and demographic shortcuts, and (iii) gradient alignment to promote device-invariant decision boundaries across clients. Built on BTS with CLAP, a multimodal language-audio pretraining model, BTS-CAFE improves the out-of-distribution ICBHI Score by 3.69 points on average over five held-out devices relative to its backbone, and outperforms conventional data augmentation and federated learning baselines in simulated device leave-out evaluations on the ICBHI and SPRSound datasets.
\end{abstract}


\section{Introduction}
\label{sec:introduction}

Respiratory sound classification (RSC) provides a non-invasive and cost-effective route for automated pulmonary disease screening and monitoring, making it attractive for telemedicine and point-of-care diagnosis~\cite{sarkar2015auscultation, bae2023patch, kim2024bts, kim2025adaptive, ge2025lungmix, koo2026empowering, kim2026quality, shridhar2026lung}. Yet clinical translation remains limited by distribution shifts across healthcare environments, under which RSC models often fail to generalize to unseen sites, acquisition conditions, or patient metadata~\cite{ge2025lungmix, koo2026empowering}.

Multi-site training is further constrained by privacy regulations and
institutional data silos~\cite{zhang2021survey}, which federated learning (FL) addresses by training a shared model without centralizing raw patient
data~\cite{mcmahan2017communication}. Recent FL work on RSC targets
heterogeneity among observed clients~\cite{kumar2025empowering, cho2026semi}, which is severe here because stethoscopes differ in frequency response, sensor sensitivity, and noise profiles~\cite{rennoll2020electronic}. Prior RSC work reduces the resulting discrepancies by adapting across observed stethoscopes~\cite{kim2024stethoscope, kim2025adaptive}; when the device is
unobserved at training time, however, adaptation no longer applies, and
device-specific style may act as a shortcut toward acquisition artifacts rather than invariant pathological markers.

Traditional data augmentation (DA) is widely adopted in RSC~\cite{yun2019cutmix, park2019specaugment, kim2024repaugment}, but exploits correlations without addressing the causal structure underlying device-induced shifts~\cite{ilse2021selecting, ouyang2022causality, jung2022understanding, koo2023comprehensive, koo2024next}. On the FL side, federated domain generalization (FedDG) methods~\cite{nguyen2022fedsr, guo2023out} learn invariant representations through regularization but ignore the data-generating process, while FedCAug~\cite{zhang2025federated} introduces causality-inspired DA into FL for generic image domains. Since respiratory recordings compose disease-relevant content and device-specific style, perturbing style while preserving content approximates an intervention on the style variable. Nevertheless, such causally motivated augmentation has not been studied for respiratory sounds.

To the best of our knowledge, this is the first work to formulate RSC as FedDG over device-associated domains, where the evaluation client corresponds to an unobserved acquisition device. Our empirical analysis shows that device style and disease content are not cleanly separable: statistical interventions reduce device information but leave residual signals, while aggressive removal harms pathology-relevant features. We therefore adapt the causality-inspired generative intervention network (GIN)~\cite{ouyang2022causality} to RSC, extending it with gain perturbation and bounded frequency-wise interpolation, and combine it with classifier-gradient alignment~\cite{guo2023out} for device-invariant decision boundaries and counterfactual metadata augmentation~\cite{koo2026empowering} against device and metadata shortcuts. Built on BTS~\cite{kim2024bts} with CLAP~\cite{wu2023large}, our proposed framework, BTS-CAFE, improves out-of-distribution (OOD) performance over conventional DA and FL baselines under simulated leave-out validation on ICBHI~\cite{rocha2017alpha} and SPRSound~\cite{zhang2022sprsound}.

\begin{figure*}[!ht]
    \centering
    \includegraphics[width=0.9\textwidth]{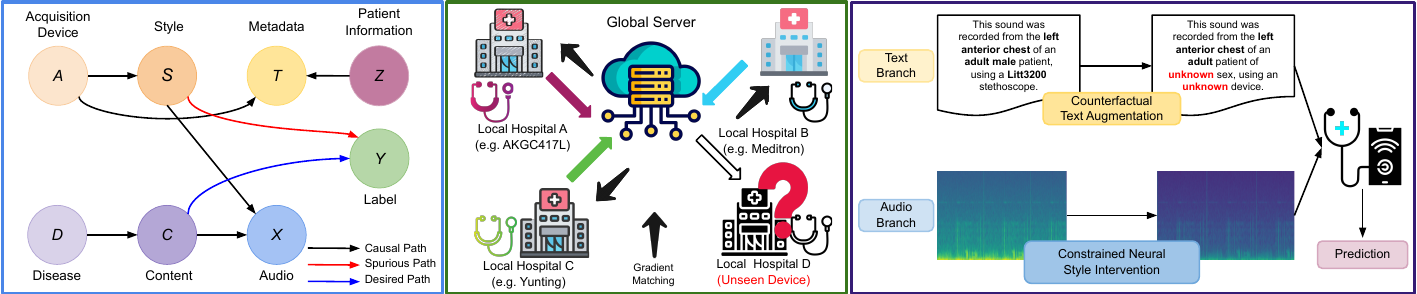}
    \caption{Overview of BTS-CAFE: \textbf{B}ridging \textbf{T}ext and \textbf{S}ound with \textbf{CA}usality-inspired \textbf{FE}derated Domain Generalization.}
    \label{fig:overview_framework}
    \vspace{-1.5mm}
\end{figure*}

\section{Preliminaries}
\label{sec:preliminaries}

\subsection{A Causal View of Respiratory Data Generation}

To motivate our augmentation design, we use a causality-inspired view of respiratory data generation. As shown in Fig.~\ref{fig:overview_framework}, an observed signal \(X\) is generated from disease-relevant content \(C\), itself determined by the underlying disease \(D\), and device-specific style \(S\), where \(C\) captures pathological patterns informative of the label \(Y\) and \(S\) captures acquisition-dependent characteristics induced by the stethoscope \(A\).

Ideally, \(Y\) should depend only on \(C\), but in practice, \(S\) can become spuriously associated with \(Y\) through site-specific acquisition protocols, acting as a shortcut. In multimodal frameworks such as CLAP~\cite{wu2023large} and BTS~\cite{kim2024bts}, text metadata \(T\) encoding device information can provide an additional shortcut path. Following causality-inspired domain generalization studies~\cite{ilse2021selecting, ouyang2022causality, jung2022understanding}, we make the following assumptions:

\begin{itemize}[leftmargin=*, itemsep=1pt, topsep=2pt, parsep=0pt]
\item \(A \rightarrow S\): the stethoscope determines device-specific acoustic style.
\item \(C, S \rightarrow X\), \(C \rightarrow Y\): the observed signal
\(X=f_X(C,S,U_X)\) entangles pathological content with device-specific style, while the label \(Y=f_Y(C,U_Y)\) depends only on content. Here, \(f_X\) and \(f_Y\) denote data-generating functions, and \(U_X\) and \(U_Y\) are exogenous variables, such as ambient noise and anatomical variations.
\item \(A, Z \rightarrow T\): the metadata prompt is determined by both device and patient information, providing an additional shortcut path through the text branch.
\end{itemize}

\subsection{An Empirical Analysis of Stethoscope-induced Bias}

\begin{figure}[ht]
    \centering
    \includegraphics[width=1.0\columnwidth]{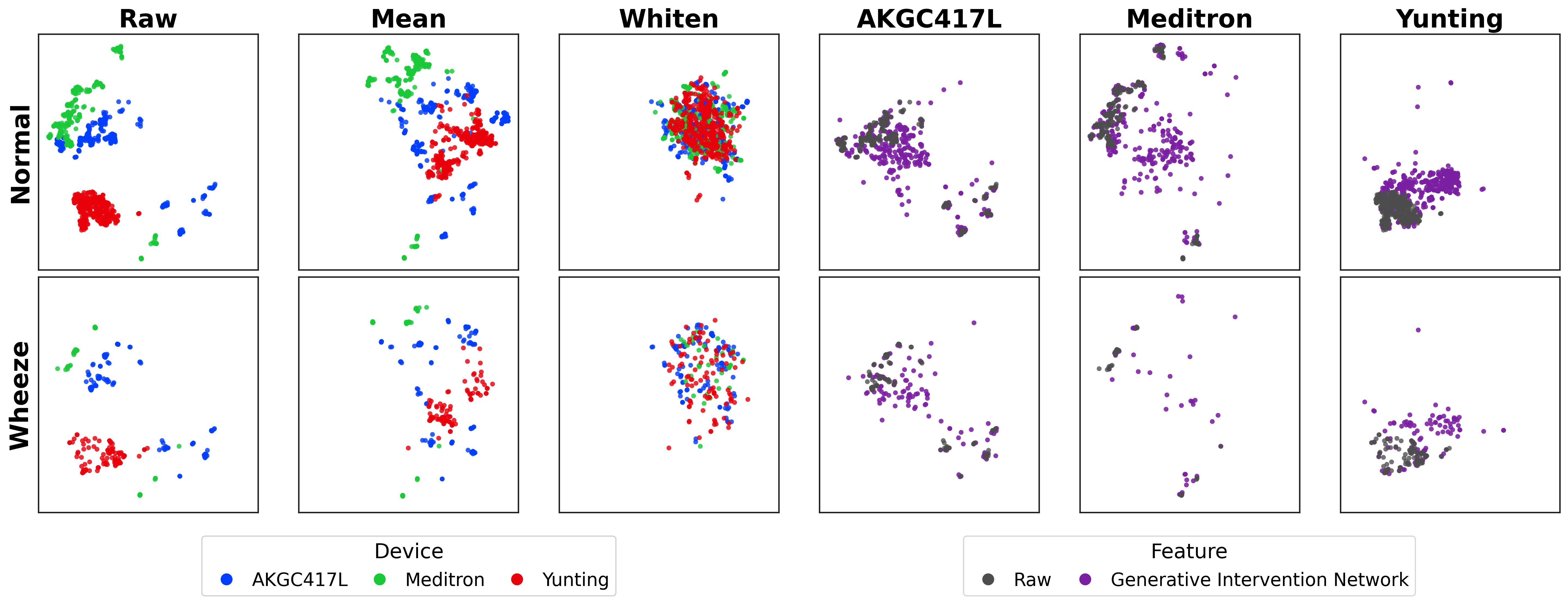}
    \caption{t-SNE of zero-shot CLAP embeddings, coloured by device. Rows: normal and wheeze. Columns: raw, background mean subtraction, low-rank whitening, and per-device views after GIN.}
    \label{fig:t_sne}
    \vspace{-1.5mm}
\end{figure}

We first examine whether acquisition devices induce shortcut cues in respiratory
sound representations, using the test sets of ICBHI~\cite{rocha2017alpha} and
SPRSound~\cite{zhang2022sprsound}, which span three heterogeneous stethoscope
groups: AKGC417L, Meditron and Yunting. In the raw space (\textbf{Raw} in
Fig.~\ref{fig:t_sne}), t-SNE of zero-shot CLAP embeddings~\cite{wu2023large}
clusters samples not only by respiratory class but also by device, suggesting
that device-specific style \(S\) is strongly encoded and may serve as a
pathology-irrelevant shortcut.

We then test whether device information can be reduced while preserving
disease-relevant content. Background mean subtraction (\textbf{Mean}) removes a device-specific mean estimated from non-event intervals, whereas low-rank whitening (\textbf{Whiten}) suppresses dominant device-related components. Under $k$-NN evaluation ($k$=50), raw embeddings yield 93.54\% device accuracy, far above the three-way chance level of 33.3\%; mean subtraction lowers this to 87.86\% and whitening to 69.53\%, still roughly twice chance, while disease accuracy stays within one point across all three settings (72.94\%, 73.01\% and 72.02\%).

These probes show that device-associated information persists well after both interventions, while disease-relevant information does not measurably improve. This motivates our stochastic spectrogram diversification (Fig.~\ref{fig:overview_framework}) for generalization to unseen device-associated domains.

\section{Methodology}
\label{sec:methodology}

\subsection{Problem Formulation}

We build on BTS~\cite{kim2024bts}, which predicts \(\hat{Y}\) from fused audio-text RSC representations. We consider an FL setting with \(K\) clients, where client \(k\) owns \(\mathcal{D}_k=\{(x_i^k,t_i^k,y_i^k)\}_{i=1}^{n_k}\sim P_k(X,T,Y)\), with \(T\) denoting the metadata-derived text prompt. Each client is associated with a recording device \(a_k\), and the observed training devices are \(\mathcal{A}_{\mathrm{train}}=\{a_k\}_{k=1}^{K}\). Client distributions differ across device partitions in acquisition conditions, patient composition, and label proportions. Our evaluation therefore measures generalization across device-associated domains.

We train a global model \(f_\theta\) using FedAvg~\cite{mcmahan2017communication}. At communication round \(r\), participating clients \(\mathcal{S}_r\) update the server model locally, and their parameters are aggregated as
\begin{equation}
    \theta^{(r+1)} =
    \sum_{k \in \mathcal{S}_r}
    \frac{n_k}{\sum_{j \in \mathcal{S}_r} n_j}
    \theta_k^{(r+1)} .
    \label{eq:fedavg}
\end{equation}

We formulate this setting as FedDG. Given \(\mathcal{A}_{\mathrm{train}}\), the model must generalize to a set of unseen devices \(\mathcal{A}_{\mathrm{test}}\)
with \(\mathcal{A}_{\mathrm{test}}\cap\mathcal{A}_{\mathrm{train}}=\emptyset\),
without accessing their data during training. The target objective is
\begin{equation}
    \min_\theta \;
    \mathbb{E}_{a\sim\mathcal{A}_{\mathrm{test}}}\,
    \mathbb{E}_{(X,T,Y)\sim P_{a}}
    \left[\mathcal{L}(f_\theta(X,T),Y)\right],
    \label{eq:loss}
\end{equation}
while relying on disease-relevant cues that are invariant to \(S\) rather than on style-dependent shortcuts. Since \(P_a\) is unavailable during training, Eq.~(\ref{eq:loss}) cannot be optimized directly; we instead approximate it by intervening on \(S\) within the observed clients.

\subsection{BTS-CAFE}
\subsubsection{Causality-inspired Generative Device Style Intervention}
\label{sec:gin}

Inspired by appearance augmentation in medical image segmentation via
GIN~\cite{ouyang2022causality}, we diversify spectrogram appearance to
discourage reliance on acquisition-specific cues. Augmentation is activated
after a warm-up of \(r_w\) rounds and applied to samples selected with
probability \(p_{\mathrm{aug}}=0.25\) in each mini-batch. We first apply a gain
intervention \(x^{\mathrm{gain}}=\rho\,x\) with \(\rho\sim\mathcal{U}(0.8,1.2)\).

GIN is implemented as two non-trainable random group convolution blocks with two
intermediate channels, re-initialized at each mini-batch, with kernels drawn
from \((1,1)\), \((1,3)\) and \((3,1)\) so that perturbations stay local in time
and frequency, yielding a style branch
\(x^{\mathrm{style}}=\mathcal{G}_\xi(x^{\mathrm{gain}})\). The two are mixed
through a mask combining an energy term and a frequency term,
\begin{equation}
\begin{aligned}
\alpha_E&=\alpha_{\min}+(1-\alpha_{\min})E,\\
\alpha_F&=\alpha_{\min}+(1-\alpha_{\min})u,\\
\alpha&=\mathrm{clip}\!\left(\alpha_E\odot\alpha_F,\;\alpha_{\min},\;1\right),
\end{aligned}
\label{eq:style_mask}
\end{equation}
where \(E\) is the per-sample min--max normalized energy of \(x^{\mathrm{gain}}\)
and \(u\) is a random gate drawn at \(1/8\) of the frequency resolution and
linearly interpolated across bins, modelling the smooth spectral response of a
stethoscope. Since \(\alpha\) is largest where \(E\) is largest, the original
signal is preserved in high-energy regions containing respiratory events while
low-energy background is perturbed more strongly, which is how content
preservation is encouraged. The augmented sample is
\begin{equation}
\begin{aligned}
\tilde{x}&=(1+\gamma E)\odot
\left(\alpha\odot x^{\mathrm{gain}}+(1-\alpha)\odot x^{\mathrm{style}}\right),\\
\tilde{x}&\leftarrow\frac{\lVert x^{\mathrm{gain}}\rVert_F}
{\lVert\tilde{x}\rVert_F+\epsilon}\,\tilde{x},
\end{aligned}
\label{eq:aug_feature}
\end{equation}
with \(\alpha_{\min}=0.25\), \(\gamma=0.15\) and a small constant \(\epsilon\);
the final step restores the magnitude of the gain-adjusted spectrogram.

\subsubsection{Counterfactual Metadata Augmentation}
Since BTS~\cite{kim2024bts} uses metadata-derived text prompts that may encode device shortcuts and over-reliance on domain-specific demographic metadata, we apply counterfactual metadata augmentation~\cite{koo2026empowering} to GIN-augmented samples during training only. For these samples, device identity is always replaced with \texttt{unknown}, while other metadata fields are independently masked with probability \(p_{\mathrm{text}}=0.25\). At test time the original prompt is used unchanged, including the true name of the held-out device.

\subsubsection{Gradient Alignment-guided Model Optimization}
Following FedIIR~\cite{guo2023out}, we align clean mini-batch gradients with respect to the trainable classifier parameters \(\theta_{\mathrm{cls}}\). Let \(h_k^{r}\) denote the reference gradient computed by client \(k\) on a
single clean mini-batch from the global model at round \(r\) and sent to the
server before local training. The server forms a global reference
\begin{equation}
\bar{g}^{\,r}=\frac{1}{|\mathcal{K}_r|}\sum_{k\in\mathcal{K}_r}h_k^{r},
\label{eq:global_ref}
\end{equation}
where \(\mathcal{K}_r\) is the set of participating clients. For each local
clean mini-batch \(\mathcal{B}_k\) we compute
\(g_k^r=\nabla_{\theta_{\mathrm{cls}}}\mathcal{L}_{\mathrm{CE}}(\mathcal{B}_k)\)
and apply the alignment penalty
\begin{equation}
\mathcal{R}_{\mathrm{align}}^{k,r}
=\frac{1}{|\theta_{\mathrm{cls}}|}
\lVert g_k^r-\bar{g}^{\,r}\rVert_2^2,
\label{eq:align}
\end{equation}
where normalizing by the number of trainable classifier parameters makes
\(\lambda\) invariant to classifier width. The global reference is treated as a constant, while local gradients retain their computation graph.

Let \(\mathcal{L}_{\mathrm{non\text{-}aug}}^{k,r}\) and
\(\mathcal{L}_{\mathrm{aug}}^{k,r}\) denote cross-entropy losses on the full original mini-batch and its partially augmented counterpart, respectively. The local objective is
\begin{equation}
\mathcal{L}_{\mathrm{local}}^{k,r}
=\mathcal{L}_{\mathrm{non\text{-}aug}}^{k,r}
+\mathbf{1}[r \ge r_w+1]\left(
\mathcal{L}_{\mathrm{aug}}^{k,r}+\lambda\,
\mathcal{R}_{\mathrm{align}}^{k,r}\right),
\label{eq:local_objective}
\end{equation}
where \(\mathbf{1}[\cdot]\) is the indicator function, \(\lambda=10^{-3}\) and \(r_w=5\). \footnote{All hyperparameters were fixed a priori and shared
across all targets, without per-target tuning.}

\section{Experiments}
\label{sec:experiments}

\subsection{Experimental Setup}
\label{sec:experimental_setup}

\noindent\textbf{Dataset.}
We use ICBHI~\cite{rocha2017alpha} (Littmann 3200, Littmann Classic~II~SE,
Meditron, AKGC417L) and SPRSound~\cite{zhang2022sprsound} (Yunting).
Table~\ref{tab:datasets} shows substantial device-level heterogeneity and label
imbalance; patients recorded with more than one stethoscope are removed, so
device partitions have disjoint subjects. Following BTS~\cite{kim2024bts} we
binarize ICBHI age into pediatric/adult groups, and following
BTS-CARD~\cite{koo2026empowering} we merge crackle-related SPRSound labels into
crackle and Stridor/Rhonchi into wheeze.

\begin{table}[ht]
\centering
\resizebox{\columnwidth}{!}{%
\begin{tabular}{ccccccc}
\toprule
\midrule
\multirow{2}{*}{\textbf{Criteria}} 
& \multirow{2}{*}{\textbf{Characteristics}} 
& \multicolumn{4}{c}{\textbf{ICBHI}} 
& \textbf{SPRSound} \\ 
\cmidrule(r){3-6} \cmidrule(l){7-7}
& & \textbf{AKGC417L} & \textbf{LittC2SE} & \textbf{Litt3200} & \textbf{Meditron} & \textbf{Yunting} \\ 
\hline
\midrule
\multirow{4}{*}{\begin{tabular}[c]{@{}l@{}}Device\\ Statistics\end{tabular}} 
& Mean of $\mu_d$ & -51.31 & -90.53 & -77.62 & -77.34 & -82.73 \\
& Mean of $s_f$ & 3275.76 & 1540.72 & 409.18 & 2487.34 & 388.91 \\
& Mean of $s_t$ & 47239.05 & 11640.13 & 1141.44 & 20810.60 & 2102.70 \\
& No. of subjects & 31 & 21 & 9 & 61 & 284 \\				

\midrule
\multirow{4}{*}{\begin{tabular}[c]{@{}l@{}}Label\\ (Ratio)\end{tabular}} 
& Normal & 1879 (45.4\%) & 672 (58.7\%) & 646 (71.3\%) & 995 (75.3\%) & 6199 (77.2\%) \\
& Crackle & 1503 (36.3\%) & 132 (11.5\%) & 48 (5.3\%) & 174 (13.2\%) & 1044 (13.0\%) \\
& Wheeze & 443 (10.7\%) & 252 (22.0\%) & 186 (20.5\%) & 133 (10.1\%) & 757 (9.4\%) \\
& Both & 315 (7.6\%) & 88 (7.7\%) & 26 (2.9\%) & 19 (1.4\%) & 31 (0.4\%) \\
\bottomrule
\end{tabular}%
}
\caption{Device-wise statistics. \(\mu_d\) is the mean of background log-mel
features (dB); \(s_f\) and \(s_t\) are the mean of the top-16 eigenvalues of the
frequency- and time-axis covariances of background segments.}
\label{tab:datasets}
\end{table}

\noindent\textbf{FedDG Settings.}
We evaluate under two simulated device-partitioned FL settings, assuming each
client uses a single stethoscope. Setting~\#1 performs leave-out validation over
AKGC417L, Meditron and Yunting; Setting~\#2 pools all Littmann samples into a
single held-out target, as both appear almost exclusively in the test portion of
the official ICBHI split. Patients recorded with more than one stethoscope are
removed, so device partitions have disjoint subjects.

IND is measured on the validation splits of the training clients, averaged
across them, and OOD on the test split of the held-out device. We optimize using
Adam with a learning rate of $5{\times}10^{-5}$ and batch size of 8, with 1
local epoch per round for 30 rounds, by which point IND Score had plateaued.
Following standard training-domain model selection~\cite{gulrajani2020search},
we select the round with the highest IND Score and report the corresponding OOD
performance, applying the same criterion to all methods; the target device is
never accessed during training or model selection.

\noindent\textbf{Baselines.}
For FL, we compare against FedAvg~\cite{mcmahan2017communication},
FedSR~\cite{nguyen2022fedsr}, FedIIR~\cite{guo2023out},
PromptFL~\cite{guo2023promptfl} and FedCAug~\cite{zhang2025federated}, covering
standard FL, FedDG, multimodal FL and causal augmentation. For DA, we compare
with Mean Subtraction and Low-rank Whitening (Sec.~\ref{sec:preliminaries}),
Gain, CutMix~\cite{yun2019cutmix}, Mixup~\cite{zhang2017mixup},
RepAugment~\cite{kim2024repaugment} and SpecAugment~\cite{park2019specaugment}.
All use BTS~\cite{kim2024bts} as the backbone under the same FedDG settings.

\noindent\textbf{Evaluation Protocol.} Following~\cite{rocha2017alpha}, we
report specificity $\mathrm{Sp}$ and sensitivity $\mathrm{Se}$---the proportions
of correctly classified normal and abnormal cases---and the ICBHI Score
\((\mathrm{Sp}+\mathrm{Se})/2\), averaged over three seeds. Standard deviation is omitted due to page limits. All experiments use
PyTorch on a single NVIDIA RTX 3090.

\begin{table*}[t]
\centering
\tiny
\setlength{\tabcolsep}{2.2pt}
\renewcommand{\arraystretch}{1.12}
\resizebox{\linewidth}{!}{%
\begin{tabular}{c c  ccc ccc  ccc ccc  ccc ccc  ccc  ccc ccc}
\toprule
\midrule
\multirow{6}{*}{\textbf{Criteria}}
& \multirow{6}{*}{\textbf{Methodologies}}
& \multicolumn{18}{c}{\textbf{Setting \#1: Leave-out over AKGC417L, Meditron, and Yunting}}
& \multicolumn{9}{c}{\textbf{Setting \#2: Littmann-family leave-out}} \\
\cmidrule(lr){3-20}
\cmidrule(lr){21-29}&
& \multicolumn{6}{c}{\textbf{AKGC417L}}
& \multicolumn{6}{c}{\textbf{Meditron}}
& \multicolumn{6}{c}{\textbf{Yunting}}
& \multicolumn{3}{c}{\multirow{2}{*}{\textbf{IND}}}
& \multicolumn{6}{c}{\textbf{OOD}} \\
\cmidrule(lr){3-8}
\cmidrule(lr){9-14}
\cmidrule(lr){15-20}
\cmidrule(lr){24-29}
&
& \multicolumn{3}{c}{\textbf{IND}}
& \multicolumn{3}{c}{\textbf{OOD}}
& \multicolumn{3}{c}{\textbf{IND}}
& \multicolumn{3}{c}{\textbf{OOD}}
& \multicolumn{3}{c}{\textbf{IND}}
& \multicolumn{3}{c}{\textbf{OOD}}
& \multicolumn{3}{c}{}
& \multicolumn{3}{c}{\textbf{LittC2SE}}
& \multicolumn{3}{c}{\textbf{Litt3200}} \\
\cmidrule(lr){3-5}
\cmidrule(lr){6-8}
\cmidrule(lr){9-11}
\cmidrule(lr){12-14}
\cmidrule(lr){15-17}
\cmidrule(lr){18-20}
\cmidrule(lr){21-23}
\cmidrule(lr){24-26}
\cmidrule(lr){27-29}
&
& $\mathrm{Sp}$ & $\mathrm{Se}$ & Score
& $\mathrm{Sp}$ & $\mathrm{Se}$ & Score
& $\mathrm{Sp}$ & $\mathrm{Se}$ & Score
& $\mathrm{Sp}$ & $\mathrm{Se}$ & Score
& $\mathrm{Sp}$ & $\mathrm{Se}$ & Score
& $\mathrm{Sp}$ & $\mathrm{Se}$ & Score
& $\mathrm{Sp}$ & $\mathrm{Se}$ & Score
& $\mathrm{Sp}$ & $\mathrm{Se}$ & Score
& $\mathrm{Sp}$ & $\mathrm{Se}$ & Score \\
\hline
\midrule

\multirow{5}{*}{FL}
& FedAvg~\cite{mcmahan2017communication}
& 95.63 & 66.61 & 81.12
& 99.23  & 1.29  & 50.26
& 88.11 & 45.59 & 66.85
& 85.23 & 19.39 & 52.31
& 79.22 & 42.64 & 60.93
& 76.30 & 48.10 & 62.20
& 88.34 & 42.80 & 65.57
& 92.10 & 30.28 & 61.19
& 14.40 & 61.82 & 38.11 \\

& FedSR~\cite{nguyen2022fedsr}
& 90.03 & 73.21 & \textbf{81.62}
& 74.89 & 17.30 & 46.10
& 91.99 & 40.15 & 66.07
& 90.16 & 15.53 & 52.85
& 80.68 & 43.04 & \underline{61.86}
& 68.27 & 46.27 & 57.27
& 89.04 & 43.25 & 66.15
& 93.15 & 34.75 & 63.95
& 11.76 & 67.69 & 39.73 \\

& FedIIR~\cite{guo2023out}
& 92.46 & 67.33 & 79.89
& 98.72 & 0.78 & 49.75
& 89.12 & 43.26 & 66.19
& 80.33 & 26.21 & 53.27
& 83.80 & 39.22 & 61.51
& 63.85 & 54.50 & 59.17
& 89.13 & 41.89 & 65.51
& 93.45 & 27.54 & 60.50
& 18.27 & 61.54 & 39.90 \\

& PromptFL~\cite{guo2023promptfl}
& 87.30 & 70.32 & 78.81
& 98.40 & 1.12 & 49.76
& 96.83 & 1.75 & 49.29
& 98.03 & 0.00 & 49.02
& 85.93 & 10.95 & 48.44
& 34.52 & 19.54 & 27.03
& 96.32 & 1.06 & 48.69
& 96.43 & 2.97 & 49.70
& 96.59 & 0.00 & \textbf{48.30} \\

& FedCAug~\cite{zhang2025federated}
& 89.25 & 71.98 & 80.62
& 98.83 & 1.34 & 50.08
& 85.54 & 50.64 & \underline{68.09}
& 74.43 & 26.21 & 50.32
& 79.12 & 38.50 & 58.81
& 46.92 & 43.19 & 45.06
& 86.85 & 43.27 & 65.06
& 92.86 & 38.56 & \underline{65.71}
& 16.41 & 66.15 & 41.28 \\

\midrule

\multirow{7}{*}{DA}
& Mean
& 89.29 & 72.50 & 80.90
& 11.49 & 53.68 & 32.59
& 85.91 & 47.35 & 66.63
& 78.36 & 29.81 & \underline{54.08}
& 82.79 & 37.88 & 60.34
& 24.62 & 53.47 & 39.04
& 87.63 & 45.45 & 66.54
& 95.83 & 33.22 & 64.53
& 3.72 & 72.31 & 38.01 \\

& Whitening
& 88.11 & 74.19 & \underline{81.15}
& 5.53 & 43.97 & 24.75
& 90.20 & 42.36 & 66.28
& 77.38 & 27.18 & 52.28
& 76.90 & 40.40 & 58.65
& 72.69 & 49.61 & 61.15
& 90.83 & 39.12 & 64.97
& 95.54 & 30.08 & 62.81
& 9.29 & 69.23 & 39.26 \\

& Gain
& 91.00 & 70.06 & 80.53
& 97.66 & 4.35 & \underline{51.01}
& 91.21 & 45.02 & \textbf{68.11}
& 88.85 & 18.45 & 53.65
& 81.31 & 41.71 & 61.51
& 54.23 & 45.76 & 49.99
& 89.57 & 43.57 & \underline{66.57}
& 96.13 & 33.05 & 64.59
& 17.34 & 65.38 & 41.36 \\

& CutMix~\cite{yun2019cutmix}
& 91.11 & 69.92 & 80.51
& 99.36 & 0.78 & 50.07
& 95.01 & 39.16 & 67.09
& 96.39 & 11.65 & 54.02
& 81.17 & 41.63 & 61.40
& 81.25 & 33.93 & 57.59
& 85.23 & 43.74 & 64.49
& 93.45 & 30.51 & 61.98
& 4.64 & 71.54 & 38.09 \\

& Mixup~\cite{zhang2017mixup}
& 90.98 & 69.98 & 80.48
& 97.45 & 3.24 & 50.34
& 88.06 & 46.68 & 67.37
& 76.39 & 29.13 & 52.76
& 89.39 & 41.24 & \textbf{65.32}
& 40.00 & 42.93 & 41.47
& 86.80 & 44.91 & 65.85
& 93.54 & 37.83 & 65.68
& 9.29 & 69.23 & 39.26 \\

& RepAugment~\cite{kim2024repaugment}
& 92.76 & 68.32 & 80.54
& 98.40 & 2.01 & 50.21
& 87.64 & 44.80 & 66.22
& 85.25 & 18.45 & 51.85
& 87.30 & 35.49 & 61.40
& 84.77 & 44.22 & \underline{64.49}
& 87.16 & 44.20 & 65.68
& 95.24 & 29.24 & 62.24
& 16.72 & 65.38 & 41.05 \\

& SpecAugment~\cite{park2019specaugment}
& 92.69 & 67.35 & 80.02
& 98.51 & 0.56 & 49.53
& 90.04 & 42.75 & 66.40
& 85.57 & 21.36 & 53.47
& 83.60 & 36.57 & 60.09
& 92.12 & 24.68 & 58.40
& 88.86 & 41.70 & 65.28
& 97.92 & 30.93 & 64.42
& 14.86 & 67.69 & 41.28 \\

\midrule

\multirow{1}{*}{}
& \textbf{BTS-CAFE (Ours)}
& 87.48 & 73.41 & 80.45
& 92.02 & 13.62 & \textbf{52.82}
& 88.02 & 46.74 & 67.38
& 85.90 & 23.30 & \textbf{54.60}
& 88.22 & 34.04 & 61.13
& 92.31 & 39.07 & \textbf{65.69}
& 87.77 & 45.97 & \textbf{66.87}
& 94.35 & 38.14 & \textbf{66.24}
& 23.22 & 63.08 & \underline{43.15} \\

\bottomrule
\end{tabular}%
}
\caption{Performance comparison under two FedDG evaluation settings. Setting \#1 reports leave-out validation over AKGC417L, Meditron, and Yunting, while \#2 reports Littmann-family leave-out evaluation. Best results are in \textbf{boldface} and the second-best results are \underline{underlined}.}
\label{tab:main_results}
\end{table*}

\begin{table}[ht]
\centering
\scriptsize
\setlength{\tabcolsep}{4pt}
\renewcommand{\arraystretch}{1.15}
\resizebox{\linewidth}{!}{%
\begin{tabular}{c c  cc  cc  cc  c  cc}
\toprule
\midrule
\multirow{4}{*}{\textbf{Criteria}} 
& \multirow{4}{*}{\textbf{Method}}
& \multicolumn{6}{c}{\textbf{Setting \#1}}
& \multicolumn{3}{c}{\textbf{Setting \#2}} \\
\cmidrule(lr){3-8}
\cmidrule(lr){9-11}
&
& \multicolumn{2}{c}{\textbf{AKGC417L}}
& \multicolumn{2}{c}{\textbf{Meditron}}
& \multicolumn{2}{c}{\textbf{Yunting}}
& \multirow{2}{*}{\textbf{IND}}
& \multicolumn{2}{c}{\textbf{OOD}} \\
\cmidrule(lr){3-4}
\cmidrule(lr){5-6}
\cmidrule(lr){7-8}
\cmidrule(lr){10-11}
&
& \textbf{IND} & \textbf{OOD}
& \textbf{IND} & \textbf{OOD}
& \textbf{IND} & \textbf{OOD}
&
& \textbf{LittC2SE} & \textbf{Litt3200} \\
\hline
\midrule

\multirow{1}{*}{Full Model}
& \textbf{BTS-CAFE (Ours)} 
& \textbf{80.45} & \textbf{52.82}
& \textbf{67.38} & \textbf{54.60}
& \textbf{61.13} & \textbf{65.69}
& \textbf{66.87}
& \textbf{66.24}
& \textbf{43.15} \\

\midrule

\multirow{3}{*}{Component}
& w/o (Gain + GIN)
& 79.92 & 48.15
& 66.71 & 50.03
& 60.85 & 60.42
& 65.94 & 62.71 & 39.84 \\

& w/o Counterfactual Metadata Augmentation
& 80.21 & 50.64
& 67.02 & 52.18
& 60.74 & 63.51
& 66.28 & 64.11 & 41.36 \\

& w/o Gradient Alignment
& 80.38 & 50.91
& 66.89 & 52.87
& 61.05 & 63.14
& 66.15 & 64.42 & 41.72 \\

\midrule

\multirow{3}{*}{GIN Design}
& Uniform interpolation~\cite{ouyang2022causality}
& 79.83 & 50.37
& 66.54 & 52.41
& 60.48 & 63.08
& 66.02 & 63.83 & 41.18 \\

& w/o frequency-wise gating ($\alpha_F{=}1$)
& 80.28 & 51.05
& 66.93 & 53.42
& 61.02 & 63.87
& 66.33 & 64.71 & 42.04 \\

& w/o gain intervention ($\rho{=}1$)
& 80.31 & 51.48
& 67.11 & 53.71
& 60.97 & 64.52
& 66.50 & 65.08 & 42.38 \\

\midrule

\multirow{1}{*}{Gradient Alignment}

& Full-batch
& 79.61 & 47.95
& 66.04 & 48.86
& 60.18 & 58.91
& 65.21 & 60.94 & 38.27 \\

\bottomrule
\end{tabular}%
}
\caption{Ablation and Design Analysis under FedDG evaluation.}
\label{tab:ablation}
\end{table}

\begin{table}[ht]
\centering
\scriptsize
\setlength{\tabcolsep}{4pt}
\renewcommand{\arraystretch}{1.15}
\resizebox{\linewidth}{!}{%
\begin{tabular}{c c  cc  cc  cc  c  cc}
\toprule
\midrule
\multirow{4}{*}{\textbf{Backbone}} 
& \multirow{4}{*}{\textbf{Method}}
& \multicolumn{6}{c}{\textbf{Setting \#1}}
& \multicolumn{3}{c}{\textbf{Setting \#2}} \\
\cmidrule(lr){3-8}
\cmidrule(lr){9-11}
&
& \multicolumn{2}{c}{\textbf{AKGC417L}}
& \multicolumn{2}{c}{\textbf{Meditron}}
& \multicolumn{2}{c}{\textbf{Yunting}}
& \multirow{2}{*}{\textbf{IND}}
& \multicolumn{2}{c}{\textbf{OOD}} \\
\cmidrule(lr){3-4}
\cmidrule(lr){5-6}
\cmidrule(lr){7-8}
\cmidrule(lr){10-11}
&
& \textbf{IND} & \textbf{OOD}
& \textbf{IND} & \textbf{OOD}
& \textbf{IND} & \textbf{OOD}
&
& \textbf{LittC2SE} & \textbf{Litt3200} \\
\hline
\midrule

\multirow{3}{*}{AST~\cite{gong2021ast}}
& AST-CE~\cite{bae2023patch}
& 80.06 & \textbf{53.99}
& \underline{71.09} & \textbf{61.88}
& 60.00 & 48.38
& \textbf{77.87} & 42.58 & 34.17 \\

& PatchMix-CL~\cite{bae2023patch}
& 79.26 & 51.32
& 64.96 & \underline{58.81}
& 54.26 & 52.88
& 65.47 & 56.58 & 35.90 \\

& SG-SCL~\cite{kim2024stethoscope}
& \textbf{81.65} & 52.10
& \textbf{71.19} & 57.38
& \textbf{62.25} & 58.64
& \underline{77.47} & 52.40 & 38.35 \\

\midrule

\multirow{3}{*}{CLAP~\cite{wu2023large}}
& BTS~\cite{kim2024bts}
& \underline{81.12} & 50.26
& 66.85 & 52.31
& 60.93 & \underline{62.20}
& 65.57 & 61.19 & 38.11 \\

& BTS-CARD~\cite{koo2026empowering}
& 80.24 & 50.18
& 65.12 & 51.47
& 60.72 & 60.38
& 65.21 & \underline{61.74} & \underline{40.92} \\

& \textbf{BTS-CAFE (Ours)}
& 80.45 & \underline{52.82}
& 67.38 & 54.60
& \underline{61.13} & \textbf{65.69}
& 66.87 & \textbf{66.24} & \textbf{43.15} \\

\bottomrule
\end{tabular}%
}
\caption{Comparative studies with existing RSC methods under FedDG evaluation.}
\label{tab:rsc_comparison}
\end{table}

\subsection{Experimental Results}
\subsubsection{Main Results}

Table~\ref{tab:main_results} reports the results under the two FedDG settings.
In Setting~\#1, BTS-CAFE achieves the highest OOD Score on all three held-out
devices, although not always the best IND performance, which is expected since
style diversification and gradient alignment trade source-domain fit for
transfer; in Setting~\#2 it attains the best Score on IND and LittC2SE, and the
second-best on Litt3200. Averaged over the five OOD targets, BTS-CAFE improves
the Score from 52.81 to 56.50 (+3.69 pp) over BTS, with the largest gains on the
jointly held-out Littmann devices. Absolute OOD performance nonetheless remains
limited: a constant-normal predictor attains Sp=100 and Se=0, giving the ICBHI
Score a label-independent reference of 50, and on AKGC417L, Meditron and the
small, highly skewed Litt3200 target all compared methods remain close to or
below it. We therefore read our results as relative robustness gains under a
shared protocol rather than as evidence of clinically deployable generalization.

\subsubsection{Ablation and Design Analysis}

\noindent Table~\ref{tab:ablation} presents ablation results. Removing our Gain+GIN
causes the largest OOD drop, indicating that style diversification is the
primary driver of cross-device generalization, while removing counterfactual metadata augmentation
or gradient alignment degrades OOD performance with IND largely stable,
suggesting complementary regularization. Disabling frequency-wise gating or gain
intervention each leads to smaller but consistent OOD drops. Lastly, estimating
each client's reference gradient from eight mini-batches instead of one lowers
mean OOD Scores across all five targets; as the single-batch estimator
additionally penalizes within-client gradient variance, the alignment term may
act partly as a variance regularizer rather than through cross-client invariance
alone, which we leave for further analysis.

\subsubsection{Comparison with Existing RSC Methods}

\noindent Table~\ref{tab:rsc_comparison} compares BTS-CAFE with existing RSC
methods under FedDG evaluation. It improves OOD performance over
BTS~\cite{kim2024bts} and BTS-CARD~\cite{koo2026empowering} on all held-out
devices, particularly on Yunting and the Littmann-family targets; BTS-CARD is
the weaker of the two, likely due to the difficulty of aggregating its multiple
components under FL. AST-based methods~\cite{bae2023patch, kim2024stethoscope}
achieve strong IND scores but exhibit larger variability across target devices.
These results indicate that BTS-CAFE improves FedDG robustness within the
same-backbone setting; cross-backbone comparisons should be read with care given
their architectural differences.

\section{Conclusion}
\label{sec:conclusion}

We presented BTS-CAFE, a FedDG framework for RSC combining causality-inspired
device style intervention, counterfactual metadata augmentation, and gradient
alignment, motivated by the partial entanglement between device style and
disease content. It yields consistent relative gains over same-backbone
baselines under two leave-out validation settings, although absolute OOD performance---especially
sensitivity---remains low across all compared methods. Future work will explore
four directions: (i) closing the remaining gap to a clinically usable operating
point, especially on small, highly imbalanced targets where all current methods
approach or fall below the reference Score of 50; (ii) extending the framework
to other audio encoders such as AST; (iii) simulating larger federations with
more clients and devices; and (iv) addressing privacy concerns arising from
gradient communication.


\newpage

\bibliographystyle{IEEEbib}

\begingroup
\footnotesize
\bibliography{refs}
\endgroup

\end{document}